\documentclass{kapproc} 
\usepackage{epsfig,bm,epsf,float,amssymb}
\usepackage{latexsym}

\let\footnote\savefootnote

\kluwerbib

\let\lcitebracket(
\let\rcitebracket)

\begin{document}

\articletitle{Atom-Optics Billiards}

\articlesubtitle{Non-linear dynamics with cold atoms in optical
traps}

\author{Ariel Kaplan, Mikkel Andersen, Nir Friedman and Nir Davidson}
\affil{Department of Physics of Complex Systems\\
Weizmann Isntitute of Science, Rehovot, Israel}
\email{akaplan@wisemail.weizmann.ac.il}


\begin{abstract}
We present a new experimental system (the ``atom-optics billiard'') and demonstrate chaotic and regular dynamics
of cold, optically trapped atoms. We show that the softness of the walls and additional optical potentials can be
used to manipulate the structure of phase space.
\end{abstract}


\section{Introduction} \label{ch:intro}

The possibility of using lasers to manipulate, cool and trap neutral atoms has revolutionized many areas of atomic
physics, and opened new research fields. In particular, cold atomic samples trapped in far-detuned optical traps
are an ideal starting point for further experimental work, since the atoms can be confined to a nearly
perturbation-free environment, Doppler shifts are almost suppressed, and long interaction times are possible
\cite{Grimm00}.

Cold atoms have been used before in non-linear dynamics experiments. First, by confining atoms to the minima of a
time-dependent amplitude (or phase) periodic optical potential, effects such as dynamical localization were
observed for ultra-cold sodium atoms \cite{Raizen99}. More recently, a similar system was used to observe
chaos-assisted tunnelling \cite{Steck01,Hensinger01}.

The ``billiard'' (a particle moving in a closed region of coordinate space and scattering elastically from the
surrounding boundary) is a well known paradigm of non-linear dynamics and much theoretical work has been done
about its properties. An interesting \emph{experimental} manifestation of classical billiard dynamics is the
observation of conductance fluctuations in semiconductor microstructures in the coherent ballistic regime
\cite{Jalabert90,Marcus92,Baranger98}. The structure of this fluctuations is affected by the classical dynamics of
the electrons. Chaotic dynamics were also exploited to construct lasers with a very high power directional
emission \cite{Gmachl98}.

We present a new experimental system, the ``atom-optics billiard'' in which a rapidly scanning and tightly focused
laser beam creates a time-averaged quasi-static potential for ultra-cold Rb atoms \cite{Friedman00}. By
controlling the deflection angles of the laser beam, we create various billiard shapes. As opposed to textbook
billiards and other experimental realizations, the particles in our system have an internal quantum structure. The
internal degrees of freedom are coupled to the external ones, and spectroscopy can serve as a sensitive probe for
the dynamics \cite{Andersen03}.

In this lectures, we review several classical non-linear dynamics experiments in atom-optics billiards. The
dynamics of the atoms confined by such billiards is studied by measuring the decay in the number of confined atoms
through a small hole on the boundary. We demonstrate the existence of regular and chaotic motion, and study the
effects of scattering by impurities on the dynamics \cite{Friedman01}. Next, we study the changes in the structure
of phase-space induced in a chaotic billiard when the billiard's boundaries are soft and find that in a chaotic
Bunimovich stadium soft walls induce islands of stability in phase-space and greatly affect the system as a whole
\cite{Kaplan01,Kaplan04}. Finally, we find that introducing additional potentials can stabilize an otherwise
chaotic billiard, or alternatively destabilize an otherwise regular one, by inducing curvature in the particles
trajectories between bounces from the billiards walls \cite{Andersen02}.

\section{Optical Dipole Traps} \label{forces}

The interaction of a neutral atom and a light field is governed by the dipole interaction, and is usually
separated into two terms which correspond to a \emph{reactive} force and a \emph{dissipative} force. When an atom
is exposed to light, the electric field component $\vec{E}$ induces a dipole moment $\vec{d}$ in the atom,
oscillating at the driving light frequency. The amplitude of the dipole moment is related to the amplitude of the
field by $d= \alpha E$, where $\alpha$ is the atomic complex polarizability, which is a function of the driving
frequency. The interaction of the induced potential with the driving field gives rise to the potential:

\begin{eqnarray}
U_{dip}=-\frac{1}{2}\left\langle\vec{d} \cdot \vec{E}\right\rangle=-\frac{1}{2 \epsilon_0 c} \mathrm{Re}(\alpha)
I(\vec{r}),
\end{eqnarray} where $I=2 \epsilon_0 c \left|E \right|^2$ is the light intensity. The reactive ``dipole'' force is a conservative one, and is equal to the gradient of the above potential. The dissipative
force is related to the power the oscillating dipole absorbs from the field, which is given by:
\begin{eqnarray}
P_{abs}=\left\langle\dot{\vec{d}} \cdot \vec{E}\right\rangle=\frac{\omega}{\epsilon_0 c} \mathrm{Im}(\alpha)
I(\vec{r}). \label{diss}\end{eqnarray}

In a quantum picture, the dipole force results from absorption of a photon from the field followed by stimulated
emission of a photon into a different mode of the laser field. The momentum transfer is the vector difference
between the momenta of the absorbed and emitted photons. The dissipative component has its origin in cycles of
absorption of photons, followed by spontaneous emission, in a random direction. Using Eq. \ref{diss} we can write
then an equation for the rate of spontaneous photon scattering:
\begin{eqnarray}
\Gamma_{scatt}=\frac{P_{abs}}{\hbar \omega}=\frac{1}{\hbar \epsilon_0 c} \mathrm{Im}(\alpha) I(\vec{r}).
\end{eqnarray}
The atomic polarizability can be calculated by using the solutions of the optical Bloch equations, while
translational degrees of freedom are taken into account \cite{Cohen-Tannoudji90}. Using this result for a
two-level atom, introducing the ``Rotating Wave Approximation'', and assuming $\delta \gg \gamma$, the dipole
potential and spontaneous photon scattering rate can be written as:
\begin{eqnarray}
U_{dip}\left( \vec{r}\right) &=&\frac{3\pi c^{2}}{2\omega _{0}^{3}}\frac{\gamma }{\delta }I\left( \vec{r}\right) \label{sdip}\\
\Gamma_{scat}\left( \vec{r}\right) &=&\frac{3\pi c^{2}}{2 \hbar \omega _{0}^{3}}\left(\frac{\gamma }{\delta
}\right)^2 I\left( \vec{r}\right)\label{sscat}
\end{eqnarray}
where $\hbar \omega_0$ is the energy separation of the atomic levels, $\delta\equiv\omega-\omega_0$ the laser
detuning and $\gamma$ the natural linewidth of the transition.

The dissipative part of the atom-light interaction is used for laser cooling of atoms \cite{Cohen-Tannoudji90}, a
pre-requisite for optical trapping. However, spontaneous photon scattering is in general detrimental to
\textit{trapped} atoms, mainly since it can induce heating and loss. Comparing the expressions for the dipole
force and scattering rate, under the above approximations, yields the relationship
\begin{eqnarray}
\frac{U_{dip}}{\hbar \Gamma _{scat}}=\frac{\delta }{\gamma },
\end{eqnarray}
which indicates that a trap with an arbitrarily small scattering rate can be achieved by increasing the detuning
while maintaining the ratio $I/\delta $.

Equations \ref{sdip} and \ref{sscat} indicate that if the laser frequency is smaller than the resonance frequency,
i.e. $\delta<0$ (``red-detuning'') the dipole potential is negative and the atoms are attracted by the light
field. The minima of the potential is found then at the position of maximum intensity. In the case $\delta>0$
(``blue detuning'') the minima of the potential is located at the minima of the light intensity.

Trapping atoms with optical dipole potentials was first proposed by Letokhov \cite{Letokhov68} and Ashkin
\cite{Ashkin70} . Chu and coworkers \cite{Chu86}, were the first to realize such a trap, trapping about 500 atoms
for several seconds using a tightly focused red-detuned beam. Later, a far-of-resonant-trap for Rb atoms was
demonstrated \cite{Miller93}, with a detuning of up to 65 nm, i.e. $\delta>5\times 10^6 \gamma$. In this case, the
potential is nearly conservative and spontaneous scattering of photons is greatly reduced. A comprehensive review
of the different schemes and applications of such optical dipole traps is presented in \cite{Grimm00}. In the
limiting case where the frequency of the trapping light is much smaller then the atomic resonance, trapping is
still possible, practically with no photon scattering \cite{Takekoshi95}. Such a quasi-electrostatic trap, formed
by two crossed CO$_2$ laser beams, was used to create a Bose-Einstein condensate without the use of magnetic traps
\cite{Barrett01}.

Apart from using far-off-resonance lasers, the interaction between the light field and the atoms can be reduced by
the use of blue-detuned traps, in which atoms are confined mostly in the dark. In the first dark optical trap cold
sodium atoms where trapped using two elliptical light sheets which where intersected and formed a ''V''
cross-section \cite{Davidson95}. Confinement was provided by gravity in the vertical direction and by the beams'
divergence in the longitudinal direction. Many other configurations where proposed since \cite{Friedman02}. Of
special importance are traps formed with a single laser beam, which are particularly simple to align, hence it is
easier to optimize their properties. The simplicity of these traps also enables easy manipulation and dynamical
control of the trapping potential, its size and its shape.

\section{Experimental Realization of Atom-Optic Billiards}
\label{sec:exp-billiard}

Friedman and coworkers \cite{Friedman00} developed a new dark optical trap for neutral atoms: the ROtating Beam
Optical Trap (ROBOT). In the ROBOT a repulsive optical potential is formed by a tightly focused blue-detuned laser
beam which is a rapidly scanned using two perpendicular acousto-optic scanners (AOS's). The instantaneous
potential is given by the dipole potential of the laser Gaussian beam:
\begin{eqnarray}
U\left( x,y,t\right) = k \frac{2P}{\pi w_{0}^{2}} \exp \left[ -2\left( \left( x-x_{0}\left( t\right) \right)
^{2}+\left( y-y_{0}\left( t\right) \right) ^{2}\right) /w_{0}^{2}\right]\label{inst},
\end{eqnarray} where $(x_{0}\left( t\right) ,y_{0}\left( t\right) )$ is the curve  along which
the center of the Gaussian laser beam scans, $w_{0}$ is the laser beam waist, $P$ is the total laser power, and
$k= \frac{3\pi c^{2}}{2\omega _{0}^{3}}\frac{\gamma }{\delta }$. When the $x$ and $y$ scanners perform a
sinusoidal and cosinusoidal scan, respectively, the time-averaged intensity in the focal plane is a ring with a
Gaussian cross-section. The radial potential barrier is then given by

\begin{eqnarray}
U=k \left( \frac{2}{\pi }\right) ^{1/2}\frac{P}{2\pi r w_{0}}\label{averaged},
\end{eqnarray}
where $r$ is the scanning radius.

Several measurements were performed to prove that the potential can be regarded as a time averaged potential (e.g.
measurements of the oscillation frequencies of atoms in the trap, and the stability as a function of the scanning
frequency).

\begin{figure}[tbp]
\begin{center}
\includegraphics[width=4in] {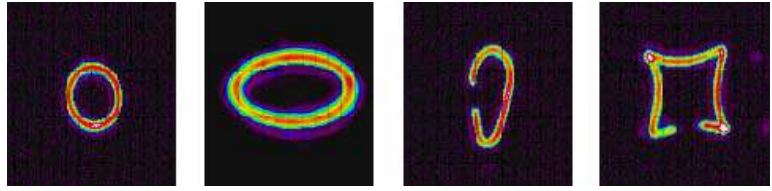} \caption{Examples of atom-optic billiards of different shapes. Shown are CCD camera images
at the rotating beam focus. From left to right: A circular billiard, an elliptical billiard, a tilted Bunimovich
stadium with a hole, and a Sinai billiard with a hole.} \label{billiard_shapes}
\end{center}
\end{figure}

We used the ROBOT to create what we denoted ``Atom-Optics Billiards'' \footnote{Optical billiards for atoms were
developed also in the Raizen group, simultaneously with our work (See \cite{Milner01})}. By controlling the
deflection angles of both AOS's using arbitrary waveform generators, arbitrary scans are performed independently
in the $x$ and $y$ direction, thus creating an arbitrary 2D scan of the beam in the focal plane. We create various
billiard shapes, such as an ellipse, a circle and a tilted Bunimovich stadium that confine the atoms in the
transverse direction (See Fig. \ref{billiard_shapes}), and probe their dynamical properties by measuring their
decay curve through a hole in their boundary.

The departure of our system from the mathematical definition of a billiard due to interactions, softness of the
walls, additional forces acting on the atoms, etc., is probably what makes it interesting, and we will devote most
of these lectures to these effects. But first, we would like to explain why, in spite of all the latter, we
\emph{can} treat our optical traps as textbook billiards.

\textbf{\emph{Three dimensional structure:}} The effects we study are two dimensional effects. Fortunately, a
large difference exists in time scales for the dynamics in the radial and longitudinal dimensions and enables us
to treat the billiards as 2D billiards, provided than our experiments are done in a short enough time.
Alternatively, to better approximate a true two-dimensional system, we confine the atoms by a stationary
blue-detuned standing wave along the optical axis. Here, the atoms are tightly confined near the node-planes of
the standing wave, but move freely in these planes, forming ``pancake'' shaped traps separated by $\sim $ 400 nm
(half the wavelength of the standing wave laser).

\textbf{\emph{Velocity distribution:}} The trap is loaded from a thermal cloud of atoms, and the trapped atoms
have a nearly thermal distribution of velocities with a root-mean-square (RMS) width of $\sim 11 v_{recoil}$, and
a typical velocity similar to this width ($v_{recoil}$ is the velocity acquired by an atoms that recoils after
absorbing a single photon, 6 mm/s for the D$_2$ line in $^{85}$Rb atoms). To better approximate a mono-energetic
case, where all atoms have identical velocity, we expose the atoms, after loading into the billiard, to a short
(1.5 $\mu $s) pulse of a strong, on-resonance, ``pushing'' beam perpendicular to the billiard beam and at 45$
^{\circ }$ to the vertical axis. Following this pushing beam, the center of the velocity distribution is shifted
to 20$v_{recoil}$, while the RMS width grows to 12$v_{recoil}$. After an additional 50 ms of collisions with the
billiard's boundaries, the direction of the transverse velocity distribution is completely randomized by the time
the hole is opened.

\textbf{\emph{Gravitation:}} Typically, the mean gravitational energy in the traps is $\frac{1}{2}mgh\leq 70
E_{recoil}$, where $h$ is the vertical size of the billiard and $E_{recoil}=\mbox{\small
$\frac{1}{2}$}mv_{recoil}^2$ is the recoil energy. After the pushing beam, the atom's kinetic energy is $\sim
400E_{recoil}$ and therefore gravity can be considered a small perturbation. In some cases, this perturbation can
be of great importance (as seen in \ref{ssec:gravity}) but in general it can be neglected.

\textbf{\emph{Collisions:}} In the range of atomic densities realized in the billiard, the mean collision time
between atoms is much longer than the experiment time, hence the motion of the atoms between reflections from the
walls can be regarded as strictly ballistic.

\textbf{\emph{Spontaneous Photon Scattering:}} Scattering events involve a change in momentum and can cause a
diffusion in phase space which is destructive to any dynamical effect. To minimize photon scattering and thus
ensure ballistic motion of the atoms and elastic reflections from the billiard potential walls, we use far-detuned
laser beams to form the billiard ($\delta=0.5-1.5$ nm $>4\cdot10^{4}\gamma$) and standing wave ($\delta=4$ nm
$>3\cdot10^{5}\gamma$).

\begin{figure}[tb]
\begin{center}
\includegraphics[width=3in] {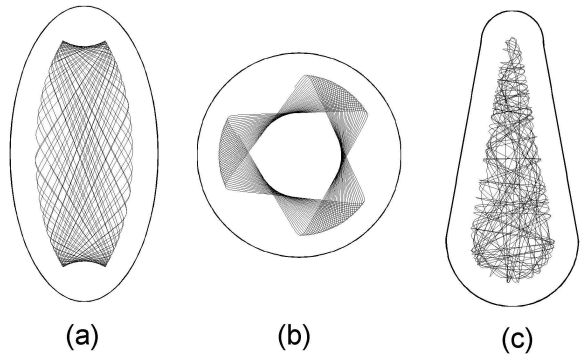} \caption{Numerical simulation of two dimensional
trajectories of Rb atoms in different atom-optics billiards. (a) Elliptical billiard: Only one of the two existing
types of trajectories is shown, for which atoms are confined by a hyperbolic caustic, and thus excluded from a
certain part of the boundary. (b) Circular billiard: Nearly periodic trajectories demand an increasingly long time
to sample all regions on the boundary. (c) ``Tilted'' stadium: Every atomic trajectory would reach a certain
region in the boundary with a comparable time scale.} \label{billiard_trajectories}
\end{center}
\end{figure}

We performed also numerical simulations in order to check the above assumption of billiard-like dynamics. The
simulations include all our experimental parameters: a Gaussian beam scanning at a finite rate, a thermal
ensemble, 3D structure of the beam and atomic cloud, the loading process. Shown in Fig.
\ref{billiard_trajectories} are typical trajectories of a single atom (the simulation shown here is, for
illustrative purposes, a 2D one). The motion of the atom shows the same features of the motion in an ideal
billiard: Periodic or quasi-periodic for the elliptical (Fig. \ref{billiard_trajectories}a) and circular (Fig.
\ref{billiard_trajectories}b) billiard and chaotic for the Bunimovich ``stadium''(Fig.
\ref{billiard_trajectories}c).

Our experimental procedure is described in \cite{Friedman01}. Briefly, $\sim 3\times 10^{8}$ atoms are loaded,
during 700 msec, from the vapor cell into a magneto-optical trap (MOT) \cite{Raab87}. Next, after 47 ms of
``weak'' and ``temporally dark'' MOT \cite{Adams95}, and 3 ms of polarization gradient cooling (PGC)
\cite{Dalibard89}, during which the trapping beam is on, we end up with $0.5-1\times$10$^{6}$ atoms in the
billiard with a temperature of $\sim 10 \mu $K and a peak density of $\sim $5$\times $10$^{10}$ cm$^{-3}$. After
all the laser beams are shut off (except for the rotating beam), a hole is opened in the billiard potential, by
switching off one of the AOS's for $\sim 1$ $\mu $s every scan cycle, synchronously with the scan. The number of
atoms remaining in the trap is measured using fluorescence detection, with a 100$\mu$s pulse resonant with
$|5S_{1/2},F=3\rangle\rightarrow|5P_{3/2},F=4\rangle$. The ratio of the number of trapped atoms with and without
the hole, as a function of time, is the main data of our experiments.

\section{Integrable and Chaotic Systems} \label{ssec:int-chaos}

The billiard is a conservative Hamiltonian system. We discuss here some general features of this class of systems.
The equations of motion for an $N$-dimensional Hamiltonian system are:
\begin{eqnarray}
\begin{array}{c @{\hspace{0.3in}} c}
\dot{q}_k=\frac{\partial\mathcal{H}(\vec{q},\vec{p})}{\partial p_k} &
\dot{p}_k=-\frac{\partial\mathcal{H}(\vec{q},\vec{p})}{\partial q_k}
\end{array}, \label{ham_eq}
\end{eqnarray}
with $k=1 \ldots N$. If there are $N$ independent, ``well-behaved'' integrals $F_i(\vec{q},\vec{p})$, that are
constant along each trajectory, then the system is called Integrable. For a conservative system, one of this
constants is the energy. The trajectory of an integrable system in phase-space lies in the intersection of
$\vec{F_i}(\vec{q},\vec{p})=f_{i}$ where $f_{i}$ are constants. Hence, the existence of $N$ integrals of motion
implies that the trajectory is confined to an $N$-dimensonal manifold, $\mathcal{M}$, in the 2$N$-dimensonal
phase-space. This $N$-dimensional manifold does not have necessarily a simple structure, but if the integrals of
motion are ``in involution'', i.e. their Poisson brackets with each other vanish $\left\{ \vec{F_j},\vec{F_k}
\right\} =0$,  then $\mathcal{M}$ has the topology of an N-dimensional torus \cite{Schuster84,Gutzwiller90}.

A natural way to describe the dynamics is by using a set of coordinates that define (i) on which torus the
trajectory takes place, and (ii) a coordinate \emph{on} the torus. The standard way to do that is by using a
canonical transformation to ``Action and Angle'' coordinates, $(\vec{I},\vec{\theta})$, which are defined as those
variables that transform $\mathcal{H}(\vec{q},\vec{p})$ into $\mathcal{H}(\vec{I})$, i.e. a Hamiltonian which does
not depend on one of the new variables. Such a transformation always exists for an integrable system
\cite{Gutzwiller90}. The equations of motion will be then:

\begin{eqnarray}
\begin{array}{c @{\hspace{0.3in}} c}
\dot{I}_k=-\frac{\partial\mathcal{H}(\vec{I})}{\partial \theta_k}=0 &
\dot{\theta}_k=\frac{\partial\mathcal{H}(\vec{I})}{\partial I_k}=\omega_k(\vec{I})
\end{array} , \label{ham_eq2}
\end{eqnarray} where $\vec{\omega}$ is a constant. The equations are then trivially integrated giving:

\begin{eqnarray}
\begin{array}{c @{\hspace{0.3in}} c}
\vec{I}(t)=\vec{I}(0) & \vec{\theta}(t)=\vec{\omega}t+\vec{\theta}(0)
\end{array}.
\end{eqnarray}
$\vec{I}$ are the actions \emph{of} the torus, and $\vec{\theta}$ the angles \emph{on} the torus. Hence,
$\vec{\omega}=\dot{\vec{\theta}}$ are the ``frequencies'' on the torus. Close (``periodic'') orbits occur only if
the frequencies have a rational ratio, i.e. $\frac{\omega_{i}}{\omega_{j}}=\frac{m}{n}$ with $m,n\in\mathbb{N}$,
for any $i,j=1 \ldots N$. For irrational frequency ratios the orbit never repeats itself but approaches any point
on the manifold infinitesimally close in the course of time (we will call this trajectories ``quasi-periodic'').

Irregular or non-integrable trajectories fill a subspace with grater dimensionality, possibly with the same
dimensionality as the whole phase space. Some of this irregular regions of phase space are ``stochastic'', meaning
that, although they are deterministic, they display an extremely sensitive dependence on the initial conditions
and are unpredictable in the long time. The field of nonlinear dynamics is rich in ways to characterize a system
which exhibits stochastic dynamics. For example, an ``ergodic'' system is one in which time averages (over a
single trajectory) are equal to ensemble averages. This implies that every single trajectory ``samples'' the
entire phase space.  The extreme case of stochasticity are K-systems or systems exhibiting ``strong chaos'', which
are defined as systems in which initially close orbits separate exponentially. Terms such as Mixing and positive
K-entropy are also commonly used. We will not be concerned with this fine-tuned hierarchy and will denote our
irregular systems, in a loose way of speaking, ``chaotic''.

It should be noted, that circular, rectangular and elliptical billiards (which exhibit integrable motion), and
Sinai and Bunimovich billiards (in which the dynamics is completely chaotic), are the exemptions in Nature. Most
systems have a ``mixed'' phase space in which regular and chaotic regions coexist.

\section{Poincar\'{e} Surface of Section}
\begin{figure}[tbp]
\begin{center}
\includegraphics[width=3.5in] {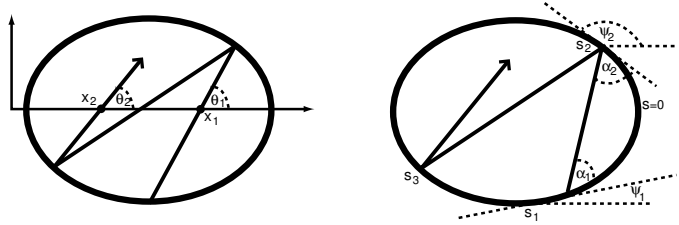}
\end{center}
\caption{Two ways to construct a Poincar\'{e} surface of section. (a) Phase space is represented using $x$ (the
points at which the orbit crosses the $y=0$ line), and $p_x=v\,\cos \theta$. (b) Bouncing Map: the usual
coordinates are the arc length, $s$, and the tangential momentum $p=\cos \alpha$.} \label{section_billiard}
\end{figure}
\label{ssec:poincare} Let us restrict ourselves to a 2-dimensional system. Phase space is four-dimensional, and
the $\mathcal{H}=E$ surface has three dimensions. A convenient way to visualize the dynamics is by using a
``Poincar\'{e} Surface of Section'' obtained by recording the successive intersections of a trajectory with an
arbitrary plane of section. For example, the $x$ surface of section $\mathcal{S}_x$ is the intersection of the
energy surface with $y=0$ and has $(p_x,x)$ as coordinates (see Fig. \ref{section_billiard}a). The value of $p_y$
is determined by the condition $\mathcal{H}(x,y=0,p_x,p_y)=E$, up to a sign. If we define it as positive, we see
that specifying the position of a system on $\mathcal{S}_x$ completely specifies its state. An initial point
$X_0=(x_0,p_{x_0})$ will cross $\mathcal{S}_x$ again at $X_1=(x_1,p_{x_1})$, $X_2=(x_2,p_{x_2})$ and so on. The
resulting map $X_i=T(X_{i-1})$ of $\mathcal{S}_x$ onto itself (the Poincar\'{e} map) is a simplified picture of
the dynamics.

Another common way to construct a Poincar\'{e} section for a billiard (see Fig. \ref{section_billiard}b), is by
looking at its bouncing map, which specifies the evolution of position and momentum from one collision with the
boundary to the next one. The position on the boundary can be parameterized either by the arc length $s$ or the
direction of the tangent, $\psi$, and the momentum can be specified using the particle's direction with respect to
the tangent, $\alpha$, or the tangential momentum, i.e. $p \equiv \cos \alpha$ (the coordinates $(s,p)$ are called
Birkhoff coordinates).

Repeated iterations of the Poincar\'{e} map reveal whether or not the motion is integrable. In the case of
integrable motion, the system explores only a 2D torus, and its intersection with $\mathcal{S}$ is a smooth,
closed curve, which becomes apparent after enough iterations. If the orbit closes (i.e. the motion is periodic)
then $X_n=X_0$ for some $n$, depending of the ratio $\frac{\omega_{1}}{\omega_{2}}$. The torus in this case is
full of such curves, and the curve will appear when the iterations are made for many different starting points
$X_0$. In the more common case of irrational $\frac{\omega_{1}}{\omega_{2}}$ the curves will be generated by
iterating any initial point $X_0$ a large enough number of iterations. For non-integrable motion tori do not
exist, and the system explores a three dimensional region. Its crossings with $S$ will not cover a curve, but a
two dimensional region of $\mathcal{S}$.

\section{Decay through a hole} \label{ssec:decay}

One way to characterize the dynamics in a billiard is by looking at the escape probability of a particle from it
through a small hole in its boundary or, alternatively, at the decay in the number of particles in it as a
function of time. This decay rate is strongly related to the decay in the system correlations, which has been a
topic of numerous studies (See for example \cite{Friedman84}). A different functional dependence is expected for
chaotic and regular billiards.

Assume that a billiard with area $A$ contains an ensemble of particles with momentum in the range $[p,p+\delta
p]$, and that there is a small hole, of length $l$, in the billiard's boundary. It can be shown \cite{Bauer90}
that if the billiard is completely ergodic, the number of particles in it, as a function of time, obeys
\begin{eqnarray}
\dot{N}(t)=N(t) / \tau_c, \hspace{0.5in} \tau_c=\frac{\pi A}{lp}.
\end{eqnarray}
Hence, the decay from a chaotic and ergodic billiard exhibits a universal (i.e. similar for all chaotic billiards)
exponential behavior, with a decay time constant $\tau_c$. The decay from an integrable billiard does not have a
universal functional dependence, and differs for different billiard shapes. Nevertheless,  power-law decays are
very common.

It is important to note already at this point that even for ``completely chaotic'' systems there can be
deviations, in the long time tail of the decay curve, from the above exponentiallity. For the Bunimovich billiard,
for example, an algebraic tail has been observed, when the hole in not infinitesimally small
\cite{Legrand90,Alt96}. This tail is induced by the marginally stable ``bouncing ball'' orbits, where a particle
bouncing back and forth between the opposite straight lines. An orbit near one of these closed orbits, will
perform a ``zig-zag'' path for many bounces (a ``resonance'') before getting lost in the chaos and the particles
can ``stick'' for a long time to the vicinity of the boundary of the marginally stable region. The phenomena of
``stickiness'' will play an important role in \ref{sec:soft-foc}. Note that in our experiments, we choose to used
a ``tilted'' Bunimovich stadium, in which the two semi-circles have different size and the straight lines are not
parallel, in order to preclude this strong position-dependent correlations between bounces \cite{Vivaldi83}.

\section{Chaotic and Integrable Dynamics}
\label{sec:chaotic-billiard} The radius of curvature of a circular billiard is constant along the boundary, and
the orbit of a particle in it is composed of a succession of chords making angles $\alpha$ with the boundary's
tangent \cite{Berry81}. The tangential momentum, $p\equiv\cos \alpha$, is then conserved and the system is
integrable. The well-known Bunimovich stadium is composed of two semicircular arcs with radius $R$ joined by
straight lines with length $L$. For any $0<L<\infty$ the system is chaotic \cite{Bunimovich74}. As already
mentioned above, we use a ``tilted'' Bunimovich stadium, in which the two semi-circles have different size and the
straight lines are not parallel.

Comparing the atomic trajectories for the circular billiard (Fig. \ref{billiard_trajectories}b) and the
tilted-stadium billiard (Fig. \ref{billiard_trajectories}c) illustrates what can be denoted a \emph{microscopic}
effect in phase space. Neither of these shapes has a \emph{macroscopic} stable region in phase space for a hole at
any point on the boundary. Nevertheless, as explained in \ref{ssec:decay}, differences in the decay rate are
expected. For the circular billiard nearly periodic trajectories exist that require an arbitrarily long time to
sample all regions on the boundary (the exactly periodic trajectories that are completely stable have only a zero
measure, and hence can be neglected). This yields many time scales for the decay through a small hole on the
boundary, and results in an algebraic decay \cite{Bauer90}. For the tilted-stadium billiard, phase space is
chaotic resulting in a pure exponential decay, with a characteristic decay time $ \tau _{c}$ \cite{Bauer90}.

\begin{figure}[tbp]
\begin{center}
\includegraphics[width=2.3in] {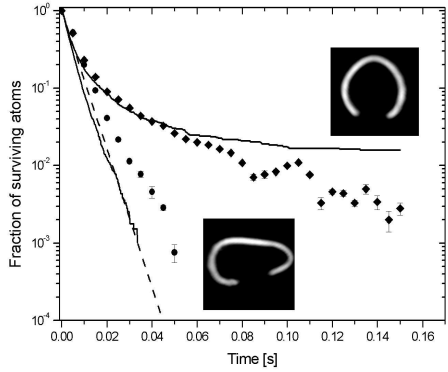} \caption{Decay of atoms from circular and stadium
billiards: The decay from the stadium billiard ($\bullet $) shows a nearly pure exponential decay. For the circle
($\blacklozenge $) the decay curve flattens, indicating the existence of nearly stable trajectories. The full
lines represent numerical simulations, including all the experimental parameters, and no fitting parameters. The
dashed line represents exp(-t/$\protect\tau _{c} $), where $\protect\tau _{c}$ is the escape time calculated for
the experimental parameters.} \label{circle_stadium_decay}
\end{center}
\end{figure}

The experimental decay curves for the circular billiard and the tilted-stadium with identical potential height,
area and hole length, and with the hole located at the bottom, are shown in Fig. \ref{circle_stadium_decay}. The
area of the circular billiard and the stadium were set to be equal, with a precision of $\sim 10\%$. The initial
decay is nearly identical for the two shapes. This is expected since it takes a certain time for the atomic
ensemble to ``feel'' the shape of the billiard. However, for longer times the decay curve for the circular
billiard flattens, indicating the existence of nearly stable trajectories, whereas that of the stadium remains a
nearly pure exponential. Also shown in Fig. \ref{circle_stadium_decay} are the results of full numerical
simulations that contain no fitting parameters. The simulations include the measured three dimensional atomic and
laser-beam distributions, atomic velocity spread, laser beam scanning, and gravity. As seen, the simulations fit
the data quite well. For long times the circular billiard simulations predict greater stability than the
experimental data. A possible explanation for this difference are imperfections of the billiard shape. Including
imperfections in the beam-shape in the simulations yielded similar effects of reducing the long-time stability of
the circular billiard. The deviation of the stadium decay curve from the simulation results, can be explained by
the existence of a small amount of atoms trapped in areas far from the focus, where the real intensity
distribution deviates from the simulated one to a greater extent. Finally, Fig. \ref{circle_stadium_decay} also
shows a pure exponential decay curve with a time constant $ \tau _{c}$ = 4.9 ms, calculated with the measured
billiard parameters. As seen, it closely resembles the full simulation results, indicating that for this
configuration, the stadium behaves nearly like an ideal chaotic billiard (two-dimensions, no gravity,
mono-energetic, zero wall thickness). Note, that deviations from a pure exponential decay, representing
correlations due to the finite hole size, are expected at long times \cite{Alt96,Vivaldi83} but occur below the
noise level of our experiment.

\section{Elliptical Billiard and the Effect of Scattering by Impurities}
\label{sec:ellipse-billiard}

The motion in an elliptical billiard is integrable (the integral of motion is the product of the angular momenta
about the two foci \cite{Berry81}). It can also be shown \cite{Koiller96} that if one segment of the trajectory
crosses the line connecting the two foci, then every segment will cross it, and hence there are two families of
trajectories: First, ``external'' trajectories are those that never cross the line between foci. They bounce all
round the billiard, and are confined outside elliptical caustics, smaller than the billiard itself but with the
same focal points. The second family, ``internal'' trajectories, are those that cross the line connecting the two
focii. They explore a restricted part of the boundary and are confined by hyperbolic caustics, again with the same
focal points. Phase space is then \emph{macroscopically} divided into two separate regions corresponding to the
types of orbits described above \cite{Koiller96}.

The decay in a system with such a non-uniform phase space, is very sensitive to the position of the hole. If it is
located at the short side of the ellipse, particles in internal trajectories will remain confined and never reach
the hole. Alternatively, all trajectories, excluding a zero-measure amount, reach the vicinity of a hole on the
long side of the ellipse and hence the number of confined particles decays indefinitely.

\begin{figure}[tbp]
\begin{center}
\includegraphics[width=2.3in] {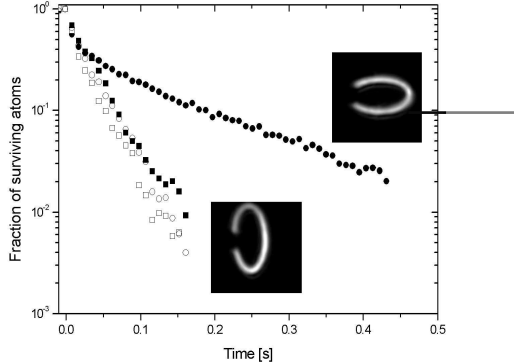} \caption{Decay of the number of atoms from an elliptical
billiard: The full symbols denote the unperturbed case, in which the surviving fraction for the ellipse with the
hole on the long side ($\blacksquare $) decays much faster than for the hole on the short side ($\bullet $). The
insets show CCD camera images of the billiards' cross-sections at the beam focus. The empty symbols show the case
in which a 10 $\protect\mu$s velocity randomizing molasses pulse is applied every 3 ms (see text). }
\label{ellipse_decay}
\end{center}
\end{figure}

Figure \ref{ellipse_decay} shows the measured survival probability for the elliptical billiard with a 60 $\mu $m
hole on the long side and on the short side. These measurements are taken without a pushing beam. To minimize the
effect of gravity, the ellipse is rotated such that the hole's direction is perpendicular to gravity for both
cases (see inset). The results reveal that the initial decay rate (for the first few points) is identical for both
cases, as expected, confirming the experimental accuracy of our shapes and holes sizes. However, at longer times,
the survival probability for the hole on the short side becomes much higher than for the hole on the long side, as
expected from the discussion above. In general, the stability due to such a macroscopic phase-space separation is
very robust and remains nearly unchanged with or without the standing wave, for a large variety of hole sizes, for
different orientations of the ellipse relative to gravity, and with small distortions of the ellipse's potential
shape.

\begin{figure}[tbp]
\begin{center}
\includegraphics[width=2.3in] {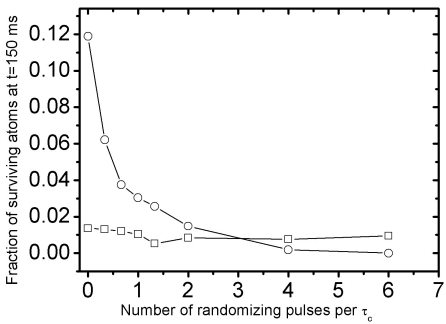} \caption{The surviving fraction of atoms 150 ms after the hole
was opened on the long ($\Box $) and short ($\circ $) side of the elliptical billiard of Fig. \ref{ellipse_decay},
as a function of the repetition rate of the velocity randomizing molasses pulses. At a rate comparable to the
escape rate, $\protect\tau_{c}^{-1}$, the stability of the trajectories was significantly reduced.}
\label{ellipse_gog}
\end{center}
\end{figure}

Next, we introduce a controlled amount of scattering to the atomic motion. We expose the confined atoms to a
series of 10 $\mu $s pulses of PGC (using the six MOT beams) every 3 ms. During each pulse, each atom scatters
$\sim $30 photons, and hence its direction of motion is completely randomized, whereas the total velocity
distribution remains statistically unchanged. Moreover, the atoms barely move during each short pulse, and
maintain ballistic motion between the pulses. Thus, the effect of the pulses resemble the effect of random
scattering from fixed points (impurities) or of short range binary atomic collisions. The measured decay curves
for this case are also shown in Fig. \ref{ellipse_gog}, for the two hole positions of the ellipse. As seen, for
the hole on the long side the randomizing molasses pulses cause little change. However, for the hole on the short
side, a complete destruction of the stability occurs, and the decay curves for the two hole positions
approximately coincide.

This effect is further illustrated in Fig. \ref{ellipse_gog}, where the measured survival probability (150 ms
after the hole was open) as a function of the repetition rate of the velocity-randomizing pulses, is shown for the
two hole positions in the ellipse. As seen, the pulse rate required to significantly reduce the stability is
approximately one pulse per $\tau _{c}$, calculated as 12 ms here, when averaging over the thermal velocity
distribution of the atoms.

\section{Billiards with Soft Walls}
\label{sec:soft-foc}

Kolmogorov-Arnold-Moser (KAM) theory provides a framework for the understanding of an integrable Hamiltonian
system which is subject to a perturbation \cite{Gutzwiller90}. As the strength of the perturbation is increased
areas of phase-space become chaotic and eventually only islands of stability, surrounded by a chaotic ``sea",
survive. The opposite question is of interest as well: What will happen to a completely chaotic system when a
perturbation is applied? will it remain chaotic or will islands of stability appear?. For the billiard system, one
such perturbation is making the walls ``soft'', as opposed to the infinitely steep potential of the ideal case.
This kind of perturbation is also interesting in the context of understanding the origins of statistical
mechanics, for which the billiard problem is a widely used paradigm \cite{Zaslavsky99}. The Sinai billiard
\cite{Sinai70} is mathematically analogous to the motion of two disks on a two-dimensional torus, an approximation
for gas molecules in a chamber, and has been proved to be ergodic. However, if the potential between the two disks
is smooth, as is the actual potential between gas molecules, it was shown that elliptic periodic orbits exist,
hence the system is not ergodic \cite{Donnay99}.


From the practical point of view, physically realizable potentials are inherently soft, and the softness of the
potential may result in a mixed phase-space with a hierarchical structure of islands. This structure greatly
affects the transport properties of the system (e.g. induces non-exponential decay of correlations), since
trajectories from the chaotic part of phase-space are trapped for long times near the boundary between regular and
chaotic motion \cite{Zaslavsky99}. As an important example, these considerations were studied in the context of
ballistic nano-structures, and found to be the cause for the fractal nature of magneto-conductance fluctuations in
quantum dots \cite{Ketzmerick96,Sachrajda98}. However, in these systems the wall softness is often accompanied by
non-ideal effects (such as scattering from impurities) and hence its role is still controversial.

For a certain kind of dispersing billiards it was theoretically proven that when the wall becomes soft, an island
appears near a singular, tangent trajectory \cite{Turaev98,Romkedar99}. More recently, and partly inspired by our
numerical results, the appearance of an additional island, this time near a corner, was also shown
\cite{Turaev03}.

\begin{figure}[tb]
\begin{center}
\includegraphics[width=4.5in] {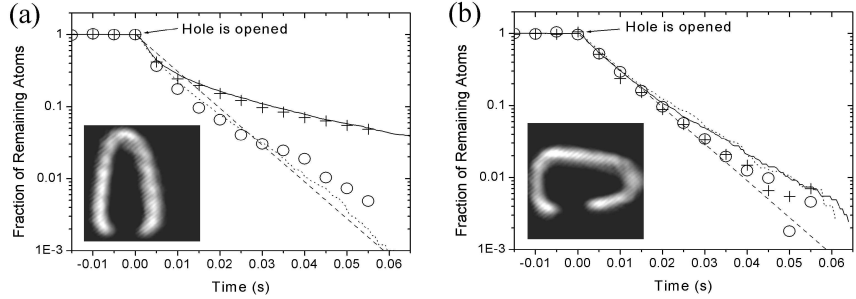}\caption{Experimental results for the decay of cold atoms
from a tilted-stadium shaped atom-optics billiard, with two different values for the softness parameter:
$w_{0}=14.5\protect\mu $m ($\circ $), and $w_{0}=24 \protect\mu $m ($+$), and for two different hole positions.
(a) The hole is located inside the big semicircle. The smoothening of the potential wall causes a growth in
stability, and a slowing down in the decay curve. (b) The hole includes the singular point, no effect for the
change in $w_{0}$ is seen. Also shown are results of full numerical simulations, with the experimental parameters
(see text) and no fitting parameters. The dashed line is $e^{-t/ \protect\tau _{c}}$, the decay curve for an ideal
(hard-wall) billiard.The insets show measured cross-sections of the (averaged) intensity of the laser creating the
soft-wall billiards, in the beam's focal plane. The size of the images is $300\times 300\protect\mu $m. }
\label{soft_stadium_decay}
\end{center}
\end{figure}

The first billiard that we study is a tilted Bunimovich stadium (see insets in Fig. \ref{soft_stadium_decay}),
composed of two semicircles of different radii ($64$ $\mu $m and $31$ $\mu $m), connected by two non-parallel
straight lines ($192$ $\mu $m long). When the potential of the wall becomes softer, a stability region appears
around the singular trajectory which connects the points where the big semicircle joins the straight lines.

We use the laser beam $1/e^2$ radius, $w_{0}$, to set the softness of the billiard's walls (See Eq. \ref{inst}),
and experimentally control it by the use of a telescope with a variable magnification, located prior to the AOS's
such that $w_{0}$ could be changed without affecting the billiard's size and shape. As seen from Eq.
\ref{averaged}, the time averaged potential is kept constant for different values of $w_{0}$ by changing the laser
power.

In Fig. \ref{soft_stadium_decay}, experimental results for the decay from a tilted stadium with two different
values of the softness parameter ($w_{0}=14.5$ $\mu $m and $ w_{0}=24$ $\mu $m), are presented. It can be seen
that when the hole is located entirely inside the big semicircle (Fig. \ref{soft_stadium_decay}a), the soft wall
causes an increased stability, and a slowing down in the decay curve. When the hole includes the singular point
where the semicircle meets the straight line (Fig. \ref{soft_stadium_decay}b), no effect for the change in $w_{0}$
is seen. We show below that these results can be explained by the formation of a stable island around the singular
trajectory, and a sticky region around it. Figure \ref{soft_stadium_decay} includes also the results of numerical
simulations, which include the three dimensional atomic and laser-beam distributions, atomic velocity spread,
laser beam scanning and gravity, and no fitting parameters. As can be seen, there is a very good agreement between
the simulated and measured decay curves. Similar decay measurements and simulations for a circular atom-optics
billiard show no dependence on $w_{0}$ in the range $14.5-24$ $ \mu $m, and no dependence on the hole position.

\begin{figure}[tbp]
\begin{center}
\includegraphics[width=4.5in] {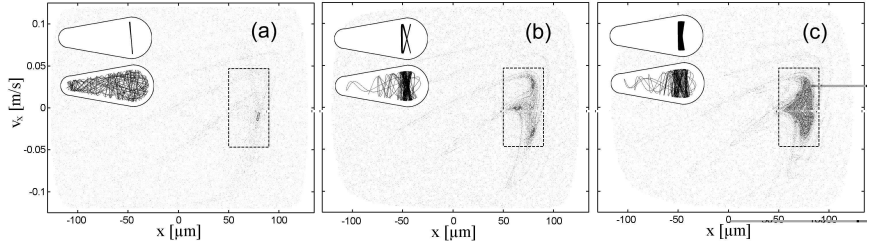}\caption{Poincar\'{e} surface of section for monoenergetic
atoms confined in a tilted-stadium atom-optics billiard with parameters specified in text, and three different
values of the softness parameter $w_{0}$. (a) $ w_{0}=18\protect\mu $m. A small elliptic island appears close to
the trajectory which connects the two singular points. Upper inset: a trajectory in the island. Lower inset: a
typical chaotic trajectory in this billiard. (b) $w_{0}=27\protect\mu $m. Three additional islands appear around
the central one, and correspond to the periodic trajectory shown in the upper inset. Around these islands there is
a large area of ''stickiness'', where the trajectories spend a long time (see trajectory in lower inset). (c) $
w_{0}=30\protect\mu $m. The three previous islands merge into one large elliptic island (see trajectory in upper
inset), with some stickiness around it (lower inset).} \label{soft_stadium_ps}
\end{center}
\end{figure}

To understand these observations, it is useful to look on how the phase-space of the system changes with changing
$w_{0}$. In Fig. \ref{soft_stadium_ps}, results of numerical simulations for classical trajectories of Rb atoms
inside the tilted-stadium billiard are shown. For clarity, we assume a monoenergetic ensemble (with
$v=20v_{recoil}$), a two-dimensional system, and no gravity. The dimensions of the billiard are equal to the
experimental ones. Phase-space information is presented using a Poincar\'{e} surface of section, showing $ v_{x}$
versus $x$ at every trajectory intersection with the billiard's symmetry axis ($y=0$), provided that $v_{y}>0$.

In Fig. \ref{soft_stadium_ps}(a), the $w_{0}=18$ $\mu $m case is shown. A small elliptic island appears around the
trajectory which connects the two singular points, as can be seen also from the upper inset, which shows a
trajectory in the island. In the lower inset, a typical chaotic trajectory is shown. For $w_{0}<12$ $\mu $m, no
islands with area larger than $10^{-4}$ of the total phase-space (the resolution of the simulations) are observed.
In general, the island size increases with the increase of $w_{0}$, as can be seen from Fig.
\ref{soft_stadium_ps}(b),(c) which correspond to $w_{0}=27,30$ $\mu $m, respectively. For $w_{0}=27$ $\mu $m,
three additional islands appear around the central one, and correspond to the periodic trajectory shown in the
upper inset. Around these islands there is a large area of ''stickiness'' \footnote{We define a ''sticky''
trajectory as one which spends inside a box surrounding the island (shown in Fig. 2) a time which is more than
$\times 3$ longer than expected for a random trajectory.}, where the trajectories spend a long time. Such a
''sticky'' trajectory is presented in the lower inset of Fig. \ref{soft_stadium_ps}(b). The exact structure of the
island and its vicinity depends on the softness parameter $w_{0}$ in a sensitive way, as can be seen from the
$w_{0}=30$ $\mu $m case (Fig. \ref{soft_stadium_ps}(c)), where the three previous islands merge with the central
one into one big elliptic island, with some stickiness around it.

\begin{figure}[tbp]
\begin{center}
\includegraphics[width=2.3in] {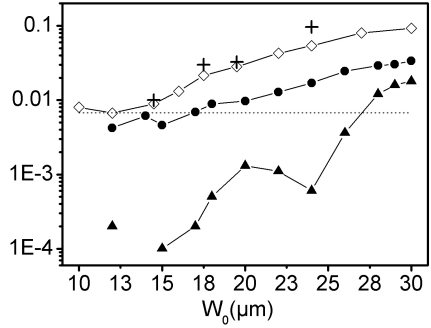}\caption{Fraction of remaining atoms at $5\protect\tau _{c}$ after the hole is open, as a function of the softness
parameter ($w_{0}$), for a hole in the big semicircle. ($+$): experimental results, ($\diamond $): numerical
simulation. The dashed line is $e^{-5}$, the expected value for small $w_{0}$. Also shown are values for the
island ($\blacktriangle $) and island + stickiness ($\bullet $) sizes as a fraction of the phase-space area,
calculated from the two-dimensional phase-space simulations.} \label{soft_stadium_gog.eps}
\end{center}
\end{figure}

In Fig. \ref{soft_stadium_gog.eps}, the measured fraction of remaining atoms at $5\tau _{c}$ ($=42.5$ ms) after a
hole is opened in the big semicircle is plotted as a function of the softness parameter, $w_{0}$, together with
the results of numerical simulations. A very good agreement exists between the decay simulations and the measured
data, and both converge to the expected value of $e^{-5}$ for small $w_{0}$. To intuitively understand the origin
of the increased stability, the results of numerical simulations, showing the relative area in phase space of the
island and the combined area of the island and the ``sticky'' regions, are also presented. The simulations reveal
that the total area size of the island and sticky regions grows monotonically with $w_{0}$, in a similar way to
the decay results. The island size itself is much smaller, and its size has a non-monotonic dependence on $w_{0}$.
These facts suggest that the remaining atoms in the experiment, at $5\tau _{c}$, are mainly due to stickiness and
demonstrate the important effect of the stickiness on the dynamics of the billiard.

\section{Billiards with Curved Trajectories}
\label{sec:curved} If particles in the billiard move in straight lines between bounces from the wall, then the
movement between them is neutral, and the dynamics is determined solely by the shape of the boundary. Exponential
separation of close trajectories (chaotic dynamics) can only occur if the boundary introduces sufficient
instability. But if particles move along curved trajectories between bounces from the wall the dynamics crucially
depends on the specific properties of this motion (we call these billiards ``curved-trajectory billiards'').
Curved trajectory billiards have been found relevant for micro devices \cite{Takagaki00}, and much effort have
been put into studying them. In these studies the curvature of the trajectory arises from a magnetic field
\cite{Tasnadi97,Robnik85,Gutkin01,Dullin98} or from the curvature of the surface on which the billiard is made
\cite{Gutkin99}. Billiards where gravity provides the confining force along its axis have been studied
experimentally, theoretically and numerically \cite{Lehtihet86,Wallis92,Hughes01,Milner01}. Theoretical work on
billiards where the curvature is caused by additional forces, both a constant force field and a parabolic
potential, have been done \cite{Dullin98}, and the stability conditions of one-periodic and symmetric two-periodic
orbits were found.

We investigate classical dynamics of curved-trajectory billiards that would have been hyperbolic if the particle
had moved in straight lines. We introduce the curvature by applying external force fields on the particle. We
investigate two special cases: (a) a uniform force field (trajectories are sections of parabolas), and (b) a
parabolic potential (trajectories are sections of ellipses if the potential is attracting, and sections of
hyperbolas if the potential is repulsive). These perturbations can lead to the formation of elliptical orbits in
hyperbolic billiards. We experimentally demonstrate these effects using atom-optics billiards.

\subsection{Periodic Orbit Stability}
\label{ssec:no-pot}

A periodic trajectory exists in the billiard if there is a ``fixed point'' in its Poincar\'{e} map, T. In general,
the map is nonlinear and cannot be written as a $2 \times 2$ matrix. In order to analyze the stability of an
orbit, we introduce the ``stability matrix'', $M$, which is a locally linearized version of $T$, and controls the
evolution of an initially infinitesimal deviation from the starting point.  Periodic orbits can be either stable
or unstable, in the sense that an initial orbit, starting at $\left(s_0+\delta s_0,p_0+\delta p_0\right)$, with
$\delta s_0$ and $\delta p_0$ infinitesimally small, will remain close to the periodic orbit starting at
$\left(s_0,p_0\right)$ or become distant from it, when the map is iterated. After $N$ iterations, the periodic
orbit returns to its initial value, and the stability of the orbit is determined by the eigenvalues of $L \equiv
M^N$, which are given by:
\begin{eqnarray}
\lambda_{\pm} = \mbox{\small $\frac{1}{2}$} \left\{ \mathrm{tr\,} L \pm \left[ \left( \mathrm{tr\,} L \right) ^2 -
4 \right] ^{1/2} \right\}.
\end{eqnarray}
If $\left| \mathrm{tr\,} L \right|<2$ then both eigenvalues have absolute value equal to unity and are complex
conjugates of the form $\lambda_{\pm}=\exp{[\pm i \beta]}$. The deviations from the periodic orbit oscillate
around zero, but remain bounded. Hence, the orbit is stable or ``elliptical''. If $\left| \mathrm{tr\,} L
\right|>2$ then both eigenvalues are real, positive and reciprocals of each other, $\left| \lambda_{\pm} \right|
=\exp{[\pm \gamma]}$. Because of the positive exponent deviations will grow exponentially and the orbit is
unstable or ``hyperbolic''.

We limit ourselves to a two-bounce orbit, with impacts on both sides at normal incidence. In this case, the trace
calculation leads to the well known geometrical stability conditions, in terms of $\rho$, the distance between the
two scattering points, and $R_{1}$ and $R_{2}$, the corresponding radii of curvature, defined as positive if the
particle scatters from a concave surface:
\begin{eqnarray}
\mathrm{tr\,} L = 4\,{\frac {{\rho}^{2}}{R_1\,R_2}}-4 \left (\frac{1}{R_2}+\frac{1}{R_1}\right )\rho+2
\label{trace}.
\end{eqnarray}
Note, that these are the same condition as geometrical optics gives for the stability of a cavity.

\subsection{A Constant Force Field}
\label{ssec:gravity} We first see how the presence of a uniform force field (constant acceleration $g$) affects
the stability. The motion in between bounces is no longer in straight lines, but along sections of a parabola. For
simplicity we assume a two-periodic orbit along the force field. We assume that the particle possesses enough
kinetic energy to access the entire billiard and further limit ourselves to an orbit which is in the direction of
the force field $\vec{F}$. In this case the linearized Poincar\'{e} map is given by:

\begin{eqnarray}
\mathrm{tr\,} L = 2-4\,\left({\frac {p_2}{R_2}}+{\frac {p_1}{R_1}}\right)\,t+4\, \frac {p_1}{R_1} \frac {p_2}{R_2}
{t}^{2}. \label{trace-g}
\end{eqnarray}
where $p_{1}$ and $p_{2}$ are the magnitudes of the velocities at encounters with the wall and t is the time it
takes the particle to move from one bounce to the next. This trace is very similar to the trace without a force
field. There are in general four solutions to the equation $\left|\mathrm{tr\,} L\right| =2$ (just as without the
force field), namely $t=0$, $t=\frac{R_{2}}{p_{2}}$, $t=\frac{R_{1}}{p_{1}}$ and
$t=\frac{R_{2}}{p_{2}}+\frac{R_{1}}{p_{1}}$. Since $\rho=-\mbox{\small $\frac{1}{2}$}\,g\,t^2+p_1t$, an
interpretation identical to the one without the force field is also valid, provided that instead of the
geometrical radii of curvature $R$, we define a pair of ``effective'' radii given by:

\begin{eqnarray}
\widetilde{R_{i}}=R_{i}+\frac{1}{2}g\frac{R_{i}^{2}}{p_{i}^{2}} \hspace{1in} i=1,2
\end{eqnarray}

Using the effective radius the stability can be found from Eq. \ref{trace}. Note that these effective radii
correspond to a new pair of center points, which are now no longer the geometrical centers but are shifted by the
force field to: $\widetilde{C_{i}}=C+\frac{1}{2} g\frac{R_{i}^{2}}{p_{i}^{2}}$. The shifted centers correspond to
points where orbits starting on the osculating circle, and perpendicular to it, will cross.

\begin{figure}[tbp]
\begin{center}
\includegraphics[width=4in]{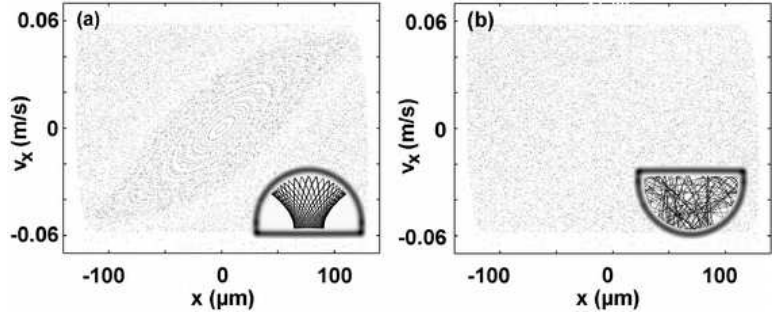}
\caption{Poincar{\'{e}} surface of sections for classical trajectories of Rb atoms inside the half Bunimovich
billiard, with the parameters used in the experiment. (a) with arc up: a large island of stability appears around
the vertical two-periodic orbit. (b) with arc down: completely chaotic phase-space. Similar phase-space diagrams
with no gravity reveal a similar completely chaotic phase-space. Insets show typical trajectories}
\label{stadium_g_ps}
\end{center}
\end{figure}

\begin{figure}[tbp]
\begin{center}
\includegraphics[width=4.5in]{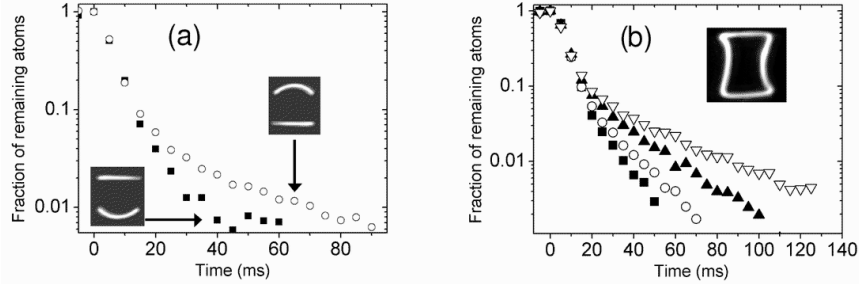}
\caption{(a) The fraction of remaining atoms as a function of time for Half a Bunimovich stadium in a constant
force field (gravity). {$\bigcirc $}: With the arc up. {$\blacksquare $}: Arc down. The existence of a stable
island in phase-space causes a slowing down in the decay. Inset: Measured light distribution of the billiard (205
{$\protect\mu m$} high), with side sections removed. (b) Fraction of remaining atoms in a dispersing billiard as a
function of time when a nearly-parabolic attractive potential is induced by a standing wave beam of variable
power. {$\blacksquare $}: 0 mW, {$\bigcirc $}: 10 mW, {$\blacktriangle $}: 20 mW and {$\triangledown $}: 30 mW.
The formation of a stable island is seen as a slowing down in the decay. Inset: The billiard. }
\label{curved_decay}
\end{center}
\end{figure}

In the shifted center picture it is now clear that a uniform force field can make stable islands in a hyperbolic
billiard of the Bunimovich kind. We demonstrate this experimentally in half a Bunimovich stadium which is obtained
by cutting it with a straight wall along its short symmetry axis (see insets in Fig. \ref{curved_decay}a). This
preserves the hyperbolicity of the billiard without a force field. As a force field we use gravity. The half
Bunimovich stadium is placed with gravity along its symmetry axis and the arc up. The expression for the
gravity-shifted centers indicates that, for atoms with low enough mechanical energy, the center of the upper arc
is shifted to outside the billiard, and thereby the orbit becomes elliptical, and an island of stability appears
(the center of the lower line is always at infinity). Alternatively if the billiard is turned upside down, i.e.
the sign of $g$ is changed, the gravity-shifted center is inside the billiard and it stays hyperbolic. We stress
that although the stability of this orbit is determined in the shifted center picture, the dynamics of the entire
billiard is not equivalent to the dynamics of a billiard without a force field constructed according to the
shifted centers. Indeed, we will see new effects such as mixed phase-space and energy dependent stability, which
would not occur in equivalent billiards without a force field.

Our half Bunimovich stadium is composed of a semicircle of radius $R_{1}=$ 158 $\mu m$ connected to a straight
section ($R_{2}=\infty $) by two 47 $\mu m$ straight sections. The velocity regime for which the vertical
two-periodic orbit is stable is 10.5 $ v_{recoil}<p_{2}<$14 $v_{recoil}$ for the arc up (velocity measured at the
lower straight section). This velocity interval is populated in the experiments due to the atomic velocity spread.
To map phase-space we performed numerical simulations of the experiment. Fig. \ref{stadium_g_ps} shows the result
of the simulations in the form of a Poincar\'{e} surface of section. As seen in Fig. \ref{stadium_g_ps}a (the arc
up), a large stable island is generated by gravity around the orbit predicted to be elliptical . We also performed
simulations without gravity, which revealed a completely chaotic phase-space, similar to the arc down picture in
Fig. \ref {stadium_g_ps}b.

The experimental results are shown in Fig. \ref{curved_decay}a. The fraction of remaining atoms is detected both
with the arc up and with the arc down (where the two-periodic orbit is always unstable). Large stability is
clearly observed for the arc up in agreement with the above description. A slowing down of the decay is seen also
with the arc down for $t>40$ $ms$. This we attribute to $\sim 1\%$ fraction of slow atoms that do not reach the
upper section of the billiard, and are trapped in the lower arc by the principle described in \cite{Wallis92}.
This is verified by observing the same fraction of remaining atoms at $t>40$ $ms$ without the upper straight
section. When the same was done with the arc up, leaving only the lower straight section, no stability was
detected, as expected.

\subsection{Parabolic Potential}
\label{ssec:parabolic} We next consider what happens if the atoms move in an attractive harmonic potential, so the
trajectories in-between scattering from the walls are sections of ellipses. For simplicity we view only what
happen to orbits along the radial direction of the potential. The motion in between bounces follows the equation
of motion $\ddot{x}=-\omega\,x$, and together with the initial conditions $x(t=0)=x_1$ and $p_x(t=0)=p_1$, we
calculate for the trace:
\begin{eqnarray}
\mathrm{tr\,} L =2\left[ 1+2\left( \frac{p_{1}p_{2}}{\omega ^{2}R_{1}R_{2}} -1\right) \sin ^{2}\omega
t-\frac{1}{\omega }\left( \frac{p_{2}}{R_{2}}+ \frac{p_{1}}{R_{1}}\right) \sin 2\omega t\right] .
\end{eqnarray}
The stability can again be found from Eq. \ref{trace} using shifted centers or effective radii. The center of the
coordinate system is at the center of the potential, and the shifted center will be at\footnote{Defining the
shifted center of the osculating circle as the point where orbits starting perpendicular to it and displaced an
infinitesimal distance along it cross, leads to that every circle have two centers (orbits are ellipses so they
cross the x-axis twice). The stability of the orbit can be determined from one of the centers, and the $\pm $
makes sure that it is the correct one that is used.}:

\begin{eqnarray}
\widetilde{C_{i}}=\pm C_{i}/\sqrt{1+\left( \omega R_{i}/p_{i}\right) ^{2}},
\end{eqnarray} where the minus refers to $R_{i}<-p_{i}^{2}/\left( \omega ^{2}\left| x_{i}\right| \right)$. $x_{i}$ is the
position of the scattering point. From the shifted centers we see that, in contrast to the uniform force field, a
two-periodic orbit can become elliptical even if $R_{1}$ and $R_{2}$ are both negative, since the $R$
corresponding to the shifted centers can be positive even though the geometrical $R$ is negative
($R_{i}<-p_{i}^{2}/ \left( \omega ^{2}\left| x_{i}\right| \right) $). This means that islands of stability can
occur around two-periodic orbits in dispersing billiards and Sinai billiards. It is also easily seen that a
potential placed asymmetric on the long symmetry axis in a Bunimovich stadium can change the stability of the
orbit.

We observed islands of stability induced by an attractive parabolic potential experimentally in a dispersing
billiard. The billiard is 220 $\mu m $\ high and consists of four convex arcs with the same curvature for each
opposite pair (see inset of Fig. \ref{curved_decay}b). Two of the arcs have a small radius of $R=-$130 $\mu $m and
the other two are very weakly curved with a radius of $R=-$1 cm. The parabolic potential is generated by
overlapping a standing wave with the billiard (an additional retro-reflected laser beam). The standing wave is
red-detuned 2.5 nm from resonance and has a radial Gaussian profile with dimensions larger than the dimensions of
the billiard ($w_{0}=244$ $\mu m$), so the potential inside the trap can be approximated with a parabolic
potential (the effect of gravity is then simply to shift the center of the potential). The decay curve was
measured for several powers of the red-detuned laser beam. The results, presented in Fig. \ref{curved_decay}b,
clearly show a slowing down of the decay as the power grows, indicating that an island of stability is formed and
grows in size for stronger attractive potentials. When the hole was placed in the bottom of the billiard, we
observed a fast decay independent of the strength of the parabolic potential, indicating that the island is indeed
formed around a vertical trajectory. It was also verified that no atoms were trapped in the red detuned standing
wave by itself for the powers we used. We performed numerical simulations that confirmed that the island is
centered around the two-periodic orbit we predict should become elliptical.

\section{Summary}

A new experimental system, the ``atom-optics billiard'', was introduced in this lectures, and several non-linear
dynamics experiments were reviewed. First, the validity of the system as an experimental realization of the
billiard model was assessed by the observation of integrable and chaotic motion of cold atoms. Next, we introduced
a controlled amount of scattering to the atomic motion by exposing the confined atoms to a series of resonant
light pulses. The measured decay curves for an originally stable case show that a complete destruction of the
stability occurs.

Next, a numerical and experimental observation of the formation of islands of stability in a billiards with soft
walls was presented. An island of stability was demonstrated in a tilted Bunimovich stadium billiard, around the
periodic trajectory that connects the points where the big semicircle joins the straight lines. Our results show
that the appearance of a stable island in a soft-wall billiard is in general accompanied by areas of
``stickiness'' surrounding it. Although the size of the stable island is a sensitive function of softness, not
always showing a monotonic behavior, the ``sticky'' trajectories modify this behavior and, in general, introduce a
monotonic slowing down in the decay of the number of trapped particles, as the walls are made softer.

Finally, we analyzed the effect of adding potentials on billiards, and found that they can cause elliptical orbits
in otherwise hyperbolic billiards. We experimentally demonstrated these effects and also found that an unstable
cavity placed in gravity can serve as a velocity selective cavity which is stable only for a narrow velocity
group.

Our research provides a well-controlled system for investigations of classical and quantum chaos, and a great
degree of control on the structure of phase space. Combined with the ability to perform precision spectroscopy,
our system may have both fundamental and practical importance: From the fundamental point of view, the
spectroscopy can serve as a very sensitive probe for the dynamics \cite{Andersen03}, and in particular shed some
light on the quantum-classical crossover \cite{Andersen04b,Andersen04c}. From the practical point of view, our
system provides a playground for studying the connection between dynamics and decoherence, which is of outmost
importance in the emerging field of Quantum computation. In addition, understanding this connection will help
develop methods to increase the coherence time for trapped atoms, and thereby further improve the precision and
accuracy of spectroscopic measurements.

\begin{acknowledgments}
The authors wish to thank U. Smilansky and V. Rom-Kedar for usefull discussions. This work was supported in part
by the Israel Science Foundation, the Minerva Foundation, and Foundation Antorchas.
\end{acknowledgments}


\end{document}